\documentclass[showpacs,amsmath,byrevtex,amssymb,superscriptaddress,aps,prb,floatfix,preprint]{revtex4}
\usepackage{graphicx}
\usepackage{hyperref}
\usepackage{color}
\usepackage{ulem}
\usepackage{dcolumn}
\begin{document}

\title{Density functional theory modeling of vortex shedding in superfluid $^4$He}

\author{Adam Freund}
\affiliation{Department of Chemistry and Biochemistry, California State University at Northridge, 18111 Nordhoff St., Northridge, CA 91330}

\author{Daniel Gonzalez}
\affiliation{Department of Chemistry and Biochemistry, California State University at Northridge, 18111 Nordhoff St., Northridge, CA 91330}

\author{Xavier Buelna}
\affiliation{Department of Chemistry and Biochemistry, California State University at Northridge, 18111 Nordhoff St., Northridge, CA 91330}

\author{Francesco Ancilotto}
\affiliation{Dipartimento di Fisica e Astronomia ``Galileo Galilei'' and CNISM, Universit\`a di Padova, via Marzolo 8, 35122 Padova, Italy and CNR-IOM Democritos, via Bonomea, 265 - 34136 Trieste, Italy}

\author{Jussi Eloranta}
\email[]{E-mail: Jussi.Eloranta@csun.edu}
\affiliation{Department of Chemistry and Biochemistry, California State University at Northridge, 18111 Nordhoff St., Northridge, CA 91330}

\date{\today}

\begin{abstract}

Formation of vortex rings around moving spherical objects in superfluid $^4$He at 0 K is modeled by time-dependent density functional theory. The simulations provide detailed information of the microscopic events that lead to vortex ring emission through characteristic observables such as liquid current circulation, drag force, and hydrodynamic mass. A series of simulations were performed to determine velocity thresholds for the onset of dissipation as a function of the sphere radius up to 1.8 nm and at external pressures of zero and 1 bar. The threshold was observed to decrease with the sphere radius and increase with pressure thus showing that the onset of dissipation does not involve roton emission events (Landau critical velocity), but rather vortex emission (Feynman critical velocity), which is also confirmed by the observed periodic response of the hydrodynamic observables as well as visualization of the liquid current circulation. An empirical model, which considers the ratio between the boundary layer
kinetic and vortex ring formation energies, is presented for extrapolating the current results to larger length scales. The calculated critical velocity value at zero pressure for a sphere that mimics an electron bubble is in good agreement with the previous experimental observations at low temperatures.
The stability of the system against symmetry breaking was linked to its ability to excite quantized Kelvin waves around the vortex rings during 
the vortex shedding process. At high vortex ring emission rates, the downstream dynamics showed complex vortex ring fission and reconnection events that appear similar to those seen in previous Gross-Pitaevskii theory-based calculations, and which mark the onset of turbulent behavior.
\end{abstract}

\maketitle

\section{Introduction}

Microscopic-level response of superfluid helium has been studied extensively by using electrons and positive ions as sensitive probes. \cite{borghesani1,williams1,thomson1,meyer1,muirhead1} In the presence of an external electric field, dissolved ions drift between the electrodes at a characteristic steady-state velocity that reflects the dissipative response of the liquid. At finite temperatures and sufficiently low electric fields, the drift velocity is observed to be directly proportional to the strength of the applied field.\cite{borghesani1} The proportionality constant is called ion mobility, which is typically determined by the viscous response of the liquid. When the steady-state ion velocity is reached, the forces due to viscous drag and the external electric field cancel. Since there is no acceleration, no dissipation due to sound emission can take place. A microscopic description of the dissipative dynamics of impurity ions drifting through liquid $^4$He at low temperatures can be obtained from state-of-the-art numerical simulations based on density functional theory (DFT) (for a recent review of DFT methods applied to superfluid $^4$He, see Ref. \onlinecite{dftreview}). Based on an extended version of this method, where viscous dissipation was added to the hydrodynamic version of the He-DFT equations, the above force balance condition has been used to compute the electron mobility in superfluid helium above 1.4 K temperature.\cite{eloranta1} The main contribution to the viscous drag was found to arise from continuous collisions between the ion and thermal rotons. The employed roton continuum approximation was observed to break down below 1.4 K where the mobility is determined by continuous interaction with thermal phonons as well as discrete roton collisions (``roton gas").

In the limit of zero temperature or at sufficiently large electric field strength, the liquid viscous response becomes negligible and the ion may no longer be able to reach the above mentioned steady-state condition. In this case, the ion will accelerate until a certain critical velocity threshold is reached after which the energy is dissipated by the creation of rotons/vorticity/turbulence.\cite{borghesani1,wilks1} Furthermore, loss of energy may also take place by emission of sound during the ion acceleration and deceleration phases, which are usually accompanied by emission of quantized vortex rings. The vortex ring emission and the possible transition to chaotic turbulent motion occurring at higher velocities has been studied previously with time-dependent Gross-Pitaevskii (GP) theory in both superfluid $^4$He and Bose-Einstein Condensates (BEC).\cite{vinen1,salman1,roberts1,sasaki1,pham1,rica1,hadzibabic1,svistunov1,tsubota1,onorato1} Although the GP equation is a rather poor model for superfluid helium, it has been shown to reproduce the vortex ring emission dynamics for an electron moving in the liquid at 0 K as well as the inherent symmetry breaking of the solution due to the emerging instability (see Ref. \onlinecite{salman1} and references therein). It is well-known, however, that GP theory can at most reproduce the phonon part of the superfluid helium dispersion relation and, consequently, it cannot provide accurate description of the vortex core structure.\cite{eloranta2} In contrast, the traditional definition of Landau critical velocity relies on the existence of roton minimum in the dispersion relation.\cite{lifshitz1,wilks1} Specifically, the flow should become dissipative when the velocity reaches the critical Landau value $v_L = \epsilon(p_{min}) / p_{min} = 59\textnormal{ m/s}$ with $\epsilon(p)$ being the superfluid dispersion relation expressed as a function of momentum, $p$. Clearly, a model that does not include description of rotons, would, within the previous reasoning, yield a critical velocity that corresponds to the speed of sound. These observations strongly suggest that the original formulation of the Landau critical velocity that considers roton emission does not directly apply to creation of vorticity. This was, in fact, first recognized by Feynman who proposed that the lowest energy excitations responsible for the onset of dissipation in superfluid helium should rather be vortices.\cite{feynman1} For a discussion on the possible microscopic-level processes that may be responsible for the existence of such critical velocity thresholds, see Ref. \onlinecite{varoquaux1} and the references therein. In order to bring the GP model to better agreement with the existing experimental electron bubble data, a modified parametrization of the GP equation has been introduced,\cite{salman1,roberts1,rica1,bowley1} at the cost of having a value for the speed of sound that is different from experiments.

Superfluid helium also serves as a unique test platform for introducing microscopic quantum corrections to classical fluid dynamics-based models. This was pioneered by Landau and Khalatnikov\cite{khalatnikov1} who introduced the famous ``two-fluid model". The two-fluid model represents the liquid in terms of the normal (viscous) and the superfluid (inviscid) liquid fractions. This model has been able to explain many unusual experimental observations such as the existence of second sound.\cite{wilks1,donnelly1} One of the current topics in this area is concerned with extending the classical Reynolds number (Re) concept to characterize the onset of vorticity and turbulence in superfluid $^4$He and BECs near 0 K (quantum regime). \cite{anderson1,schoepe1} In classical liquids, the Reynolds number diagnostics can provide an estimate for the onset of vorticity and turbulence based on the flow velocity ($v$), object size ($D$), liquid density ($\rho$), and liquid viscosity ($\eta$). However, in the absence of viscosity, this concept becomes ill-defined as $\textnormal{Re} = \frac{\rho Dv}{\eta} \rightarrow\infty$. In the spirit of the original definition of Reynolds number, it has been recently proposed that its superfluid counterpart can be obtained by replacing kinematic viscosity, $\nu = \eta/\rho$, with quantized circulation, $\Gamma = h/m_{He}$, yielding Re$_s \sim m_{He}vD/h$, where $h$ is the Planck constant and $m_{He}$ is the helium atom mass.\cite{anderson1,schoepe1} This model has been employed to analyze, e.g., oscillating sphere data in superfluid $^4$He in the mK regime,\cite{reeves,schoepe2,niemetz} where critical Re$_s$ value for the appearance of turbulent behavior was determined. Note that the existence of such a threshold value for Re$_s$ implies that the associated critical velocity for the onset of dissipation must scale as $1/D$.

After reaching the critical velocity for vortex ring shedding, the associated emission frequency is expected to increase with the flow velocity. Eventually, the system develops complex vortex tangles in the wake of the moving object (i.e., turbulence) where concerted vortex ring size reduction and proliferation take place through reconnection events between crossing vortex lines. Theoretical modeling of turbulence is especially challenging as the resulting dynamics tends to span multiple length scales. In classical liquids, a characteristic feature of turbulence is the appearance of so-called Kolmogorov $k^{-5/3}$ spectrum, which results from the breakdown of vortex rings (Richardson cascade).\cite{lifshitz1,vinen1} Although quantum turbulence appears to be  similar to its classical counterpart, some important differences are expected due to the capability of superfluid helium to sustain quantized vorticity. The general features of quantum turbulence have been reviewed elsewhere in the literature.\cite{vinen1,sreenivasan1} 

In this paper, we apply time-dependent DFT method\cite{dftreview} to model superfluid $^4$He flow past spherical heliophobic objects (``bubbles"). In addition to determining the critical velocities for vortex ring emission as a function of the sphere radius and the external pressure, we also briefly characterize the main features of the resulting liquid dynamics (e.g., symmetry breaking). To rationalize the results obtained from the simulations, we show that the onset of vortex ring shedding can be predicted by comparing the energy required to create a vortex ring around the bubble equator and the kinetic energy stored within the boundary layer in front of the bubble. This model may be applied to estimate critical velocities for objects that would otherwise be too large for microscopic calculations. Finally, the present results are compared with previous GP theory-based calculations for the electron bubble\cite{salman1} and the differences between the DFT and GP models are discussed.

\section{Theory}

\begin{figure}
\includegraphics[scale=.5]{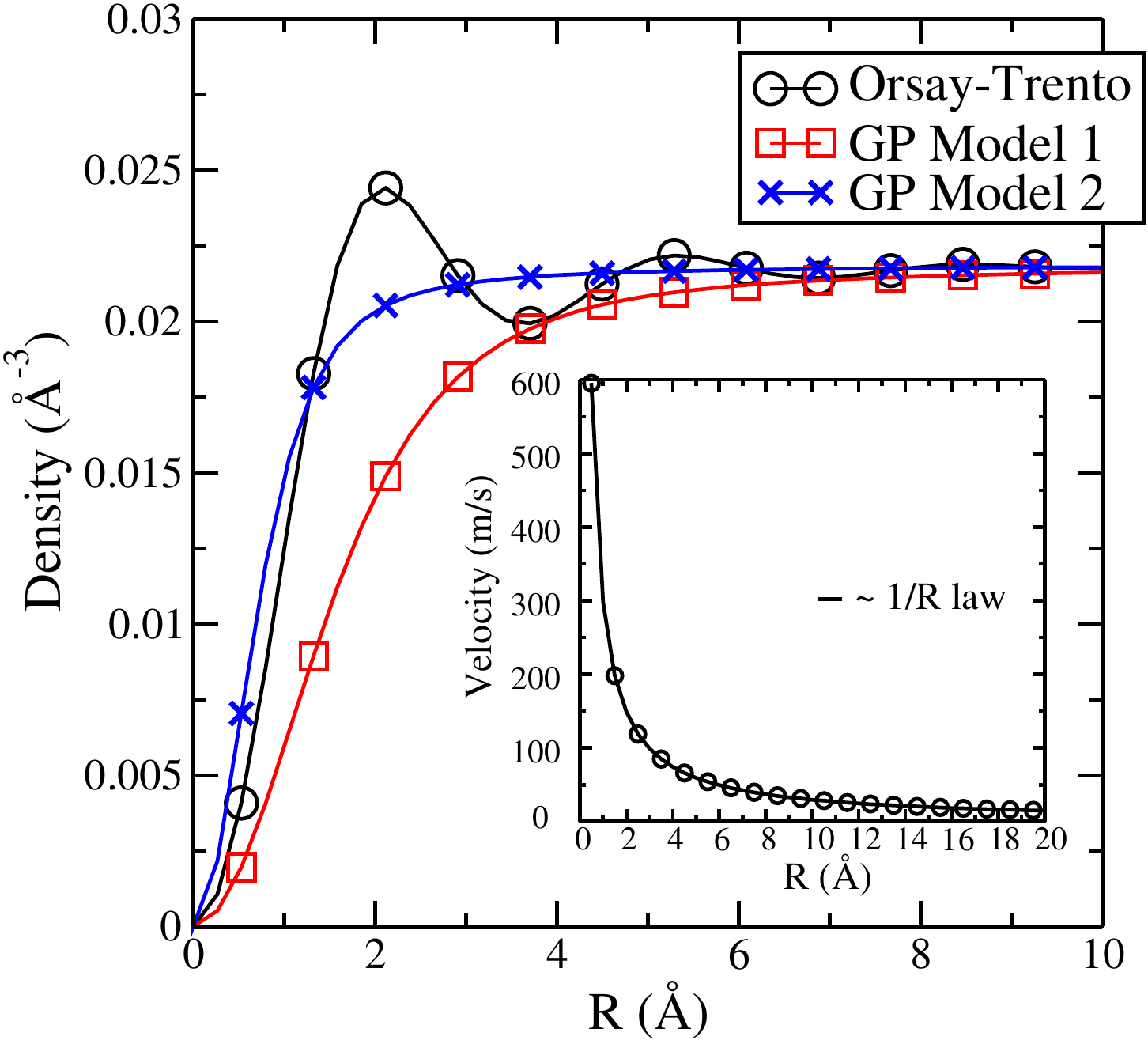}
\caption{Vortex core structure obtained using Orsay-Trento DFT (basic form without kinetic energy correlation and backflow) and the two Gross-Pitaevskii models (GP Model 1 and GP Model 2). See text for definitions of the models. The inset demonstrates that the velocity profile around a linear vortex line from OT-DFT is proportional to $1/R$ (irrotational vortex).}
\label{fig1}
\end{figure}

We model superfluid $^4$He at 0 K by time-dependent DFT (TDDFT).\cite{dftreview} Within this approach, the system is described by an energy density functional, which includes both finite-range and non-local terms that are required to describe the $T=0$ response of liquid $^4$He on the \AA{}ngstr\"om-scale accurately. The minimization of such functional results in a non-linear time-dependent Schr\"odinger equation:
\begin{equation}
i\hbar\frac{\partial}{\partial t}\psi(r,t) = -\frac{\hbar^2}{2m_{\textnormal{He}}}\Delta \psi(r,t) + \frac{\delta E_c}{\delta \rho}\psi(r,t) - \vec{v}\cdot\vec{p}\psi(r,t)
 - \mu\psi(r,t) - \frac{1}{2}m_{He}\left|\vec{v}\right|^2\psi(r,t)
\label{eq1}
\end{equation}
where $\psi(r,t)$ is the time-dependent order parameter, $m_{\textnormal{He}}$ is the helium atom mass, one particle density is obtained from $\rho = \left|\psi(r,t)\right|^2$ (in unit of atoms per unit volume), $\delta E_c/\delta \rho$ is the functional derivative of the 
correlation energy functional in the so-called 
Orsay-Trento (OT) formulation,\cite{dalfovo,dftreview,eloranta3} $\mu$ is the chemical potential, and the term containing the liquid momentum operator $\vec{p}$ introduces the flowing liquid background with velocity field $\vec{v}$ (i.e., the bubble is stationary and the liquid flows past it). The magnitude of $\vec{v}$ must be chosen such that it is compatible with the simulation box length in the direction of the flow: $\left|\vec{v}\right| = 2\pi\hbar n / \left(m_{He} L\right)$ where $n$ is an integer and $L$ is the box length. In this work, this requirement limits the accuracy of determining the critical velocity for vortex ring emission by approximately $\pm 1$ m/s. The last term in Eq. (\ref{eq1}) must be included in order to match the chemical potential of the moving bulk liquid. Note that the constraint in Eq. (\ref{eq1}) imposes constant velocity rather than constant force employed in previous calculations.\cite{salman1,eloranta1} The latter case would apply, for example, to modeling ion mobilities in the presence of an external electric field whereas our present aim is to characterize the liquid flow as a function of velocity and other system parameters. The GP theory can be obtained as a special case of Eq. (\ref{eq1}) by setting $\delta E_c/\delta\rho = \frac{\mu}{\rho_0}\left|\psi(r,t)\right|^2$ where $\rho_0$ is the bulk liquid density. In the following, we will refer to two different parametrizations of GP as Model 1: $\mu/\rho_0 = 1249.6$ K\AA{}$^{-3}$ and Model 2: $\mu/\rho_0 = 277.66$ K\AA{}$^{-3}$. Model 1 yields the correct speed of sound (230 m/s) whereas Model 2, which yields only 158 m/s, was introduced to match the correct bubble sizes as found in earlier ion solvation studies.\cite{roberts2,salman1} A comparison of the core structure for a linear vortex line obtained by OT-DFT and the two GP models is shown in Fig. \ref{fig1}. Note that OT-DFT reproduces the known damped density modulations around the vortex core whereas both GP models lack this structure due to the missing roton branch. These modulations can be viewed as a cloud of virtual roton excitations, which are sustained by the phase of the vortex wave function.\cite{reatto} Furthermore, it has been suggested that these virtual rotons may be converted to real rotons during vortex reconnection events, thus making vortex tangles a source of non-thermal rotons.\cite{reatto} 

\begin{figure}
\includegraphics[scale=.5]{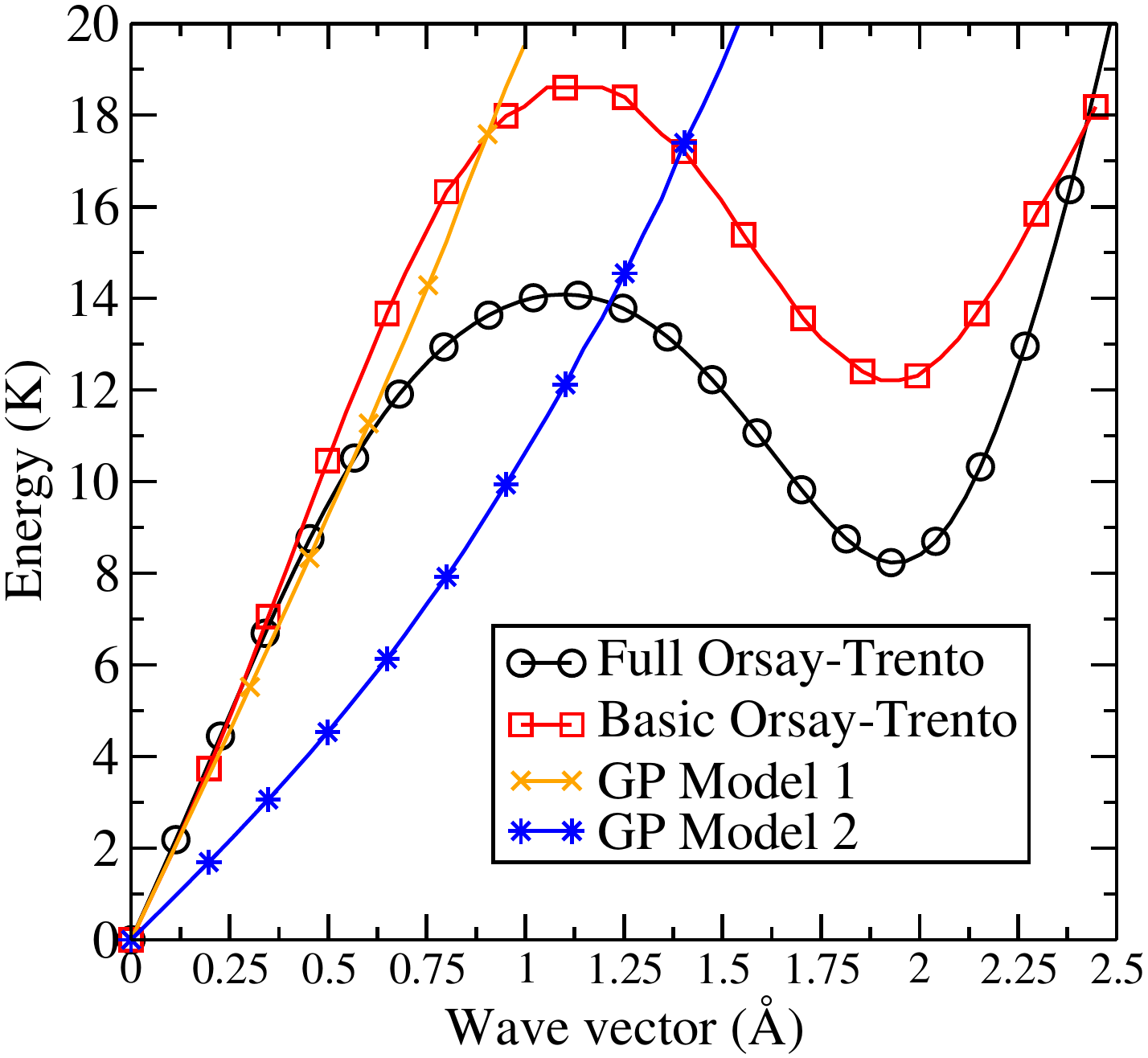}
\caption{Bulk liquid dispersion relations for full Orsay-Trento DFT (including kinetic energy correlation and backflow), the basic Orsay-Trento DFT (employed in this work), and the two Gross-Pitaevskii models (GP Model 1 and GP Model 2). See text for definitions of the models.}
\label{fig2}
\end{figure}

In this work, we employ the basic form of OT-DFT for $E_c$, which excludes the so-called backflow (BF) and non-local kinetic energy correlation (KC) functionals (for explanation of these terms, see Ref. \onlinecite{dftreview}). If these terms were included, OT-DFT would reproduce the experimental dispersion relation exactly with the exception of the turn-over region at high momenta beyond rotons.\cite{eloranta2,dftreview} The inclusion of BF and KC terms is not only computationally very expensive, but their proper numerical evaluation requires a very fine spatial grid (grid step less than 0.5 \AA{}). Given the length scales required in the present simulations, this was not possible with the available computational resources. A comparison between the bulk liquid dispersion relations produced by full OT-DFT, basic OT-DFT, and the two different GP parametrizations (Models 1 and 2) is shown in Fig. \ref{fig2}. At best, the GP theory can describe the phonon branch, but clearly it does not include the roton branch. Note also that GP Model 2 yields a much softer dispersion relation at small wave vectors and, consequently, it yields the wrong speed of sound. However, as shown in Fig. \ref{fig1}, it does yield a vortex core width that is in better agreement with the OT-DFT results than GP Model 1.

The OT-DFT model is implemented in the libdft library,\cite{libdft} which relies on libgrid\cite{libgrid} for 3-D Cartesian grid primitives. The latter library includes OpenMP and CUDA directives to achieve parallel execution on both shared memory central processing unit (CPU) and graphics processing unit (GPU) systems. We have implemented a priority-based memory management algorithm that allows hybrid CPU/GPU execution of the grid primitives by automatically synchronizing memory blocks between the host and GPU memories as needed. It is important to minimize the CPU-GPU memory transfer operations because they are very slow in comparison with any other GPU related operation. The current OT-DFT calculations employed 512x256x256 Cartesian 3-D grids with a spatial step size of 1.1 \AA{} (2.0 Bohr). Although the spatial grid step is comparable to the actual vortex core parameter (\textit{ca.} 0.79 \AA{}),\cite{eloranta2} we have verified that the current results are close to those obtained using a finer grid (down to 0.5 \AA{}), but with a smaller spatial extent of the simulation box. All calculations were carried out with double precision floating point accuracy because single precision was not sufficiently accurate for the long propagation times required in this work. Two different propagation schemes for the kinetic energy term in Eq. (\ref{eq1}) were tested: 1) direct Fast Fourier Transform (FFT)-based propagation in the reciprocal (momentum) space\cite{fftw,cufft} and 2) Crank-Nicolson (CN) method where the $x$, $y$, and $z$ directions where isolated by the operator splitting method.\cite{eloranta4,numrep} In the latter case, the time propagation step is reduced to solving a tridiagonal matrix equation (Thomas algorithm) when Neumann boundaries are imposed. While the FFT method was significantly faster than CN, it is not straight forward to implement the absorbing boundaries in the reciprocal space. The absorbing boundaries around the edges of the box for CN were implemented by gradually switching to imaginary time propagation within the buffer zone (60.0 $\times$ 25.0 $\times$ 25.0 Bohr$^3$).\cite{mateo1} However, this zone was not able to fully prevent back reflections from the boundaries (e.g., long wavelength phonons) and therefore the calculations tended to exhibit numerical artifacts at long simulation times when these reflected waves reach the object again. For this reason, all production runs employed the FFT-based propagation method without any absorbing boundaries in order to speed up the calculations. The CPU-based simulations were carried out on a 64 core AMD Opteron Linux system and GPU-based simulations on NVIDIA Titan Black (Kepler architecture with 2880 cores and 6 GB of memory) and Titan X (Pascal architecture with 3584 cores and 12 GB of memory) using the CUDA library.

The initial condition for the time-dependent OT-DFT simulations was obtained by performing 
preliminary iterations for 400 ps ("warm-up" period) where imaginary time was linearly transformed into real time propagation (constant time step magnitude 15 fs). A typical length for the real time simulations was 1.3 ns, which allowed for the observation of several vortex ring emission cycles and to determine the critical velocity threshold for vortex emission. 

All heliophobic bubbles considered in the simulations were modeled by a spherically symmetric exponentially repulsive potential of the form (``rigid" bubble):
\begin{equation}
V(r) = V_0 e^{-a_1 (r - r_m)} 
\label{eq0}
\end{equation}
\noindent
where $V_0 = 3.8003\times 10^5$ K, $a_1 = 1.6245$ \AA{}$^{-1}$, and $r_m$ is varied to achieve the desired bubble size. To alleviate numerical noise originating from mapping this potential on the relatively sparse spatial grid, a three point average was employed inside every grid step. For example, setting $r_m = 10.05$ \AA{} produces a cavity void of $^4$He atoms with a radius of \textit{ca.} 18.5 \AA{} that corresponds roughly to an electron bubble at zero pressure.\cite{pi1,eloranta5} However, this rigid repulsive potential will clearly not be able to reproduce the compressibility of real electron bubbles at higher pressures. Moreover, the rigid bubble model is expected to predict a critical velocity for vortex ring emission, which is slightly higher than the one predicted for a deformable non-spherical bubble: as the flow velocity increases, a compressible electron bubble may become squeezed along the direction of motion while it expands in the transverse directions.\cite{anci1,guo,salman1} As a consequence of this change of shape, vortical fluid motion develops around the bubble equator, which promotes the formation and emission of quantized vortex rings. 

Since the velocity profile around a vortex line in superfluid helium is inversely proportional to the distance from the vortex center (i.e., $1/R$ as demonstrated in the inset of Fig. \ref{fig1}), the vortex line is irrotational, $\vec{\nabla}\times\vec{v} = 0$. For this reason, $\vec{\nabla}\times$ operator cannot extract vorticity from the liquid velocity field. Instead, we apply this operator on the liquid current density, $\rho \vec{v}$:
\begin{equation}
c_n(\vec{r},t) = \left|\vec{\nabla}\times\left(\rho(\vec{r})\vec{v}(\vec{r})\right)\right|^n
\label{eq-circ}
\end{equation}
where $n$ is a fixed positive integer. By expressing the $^4$He order parameter in the Madelung form: $\psi(r,t) = \sqrt{\rho(r,t)}\exp\left(iS/\hbar\right)$, the liquid velocity field can be obtained from the associated phase factor as $\vec{v} = \vec{\nabla}S/m_{\textnormal{He}}$. A value of $n = 2$ in Eq. (\ref{eq-circ}) works very well for highlighting the vorticity inside the simulation box (e.g., the Volume representation in Paraview program\cite{paraview}) and is used here throughout. The total amount of vorticity, $c_{tot,n}(t)$, created up to time $t$ can be obtained by integrating Eq. (\ref{eq-circ}) over the simulation box volume:
\begin{equation}
c_{tot,n}(t) = \int c_n\left(\vec{r},t\right)d^3r
\label{eq-circ2}
\end{equation}
This procedure can identify vortex ring emission events as well as yield the total number of vortex rings emitted if the increase in this quantity is known for a single vortex ring \textit{a priori} and the possible vortex-vortex interactions and symmetry breaking effects can be neglected.

The drag force on the bubble can be evaluated by two independent equivalent expressions; first from the bubble-helium pair interaction, $V(r)$:\cite{eloranta1}
\begin{equation}
\vec{F}_{drag} = \left.-\int\rho(r',t)\vec{\nabla}_rV\left(\left|r - r'\right|\right)d^3r'\right|_{r = 0}
\label{eq0a}
\end{equation}
and, alternatively, from the rate of momentum transfer to the liquid:\cite{pi2}
\begin{equation}
\vec{F}_{drag}(t) = \left.m_{\textnormal{He}}\frac{\partial}{\partial t'}\int\rho(r,t')\vec{v}(r,t')d^3r\right|_{t' = t}
\label{eq0b}
\end{equation}
Note that Eq. (\ref{eq0b}) would fail if liquid excitations reach the absorbing boundary region of the simulation box (CN method). Both expressions were observed to produce identical results within the numerical accuracy of the computation. Another quantity related to the drag force, the hydrodynamic added mass (in units of He atoms), can be evaluated from the velocity field through:\cite{eloranta6}
\begin{equation}
M_{add}(t) = \frac{1}{v_0}\int\rho(r,t)v_x(r,t)d^3r
\label{eq0c}
\end{equation}
where the liquid flow is oriented along the $x$-axis and $v_0 = \left|\vec{v}\right|$ is the moving background velocity (see Eq. (\ref{eq1})).

\section{Results}

\begin{figure}
\vspace*{0.3cm}
\includegraphics[scale=.5]{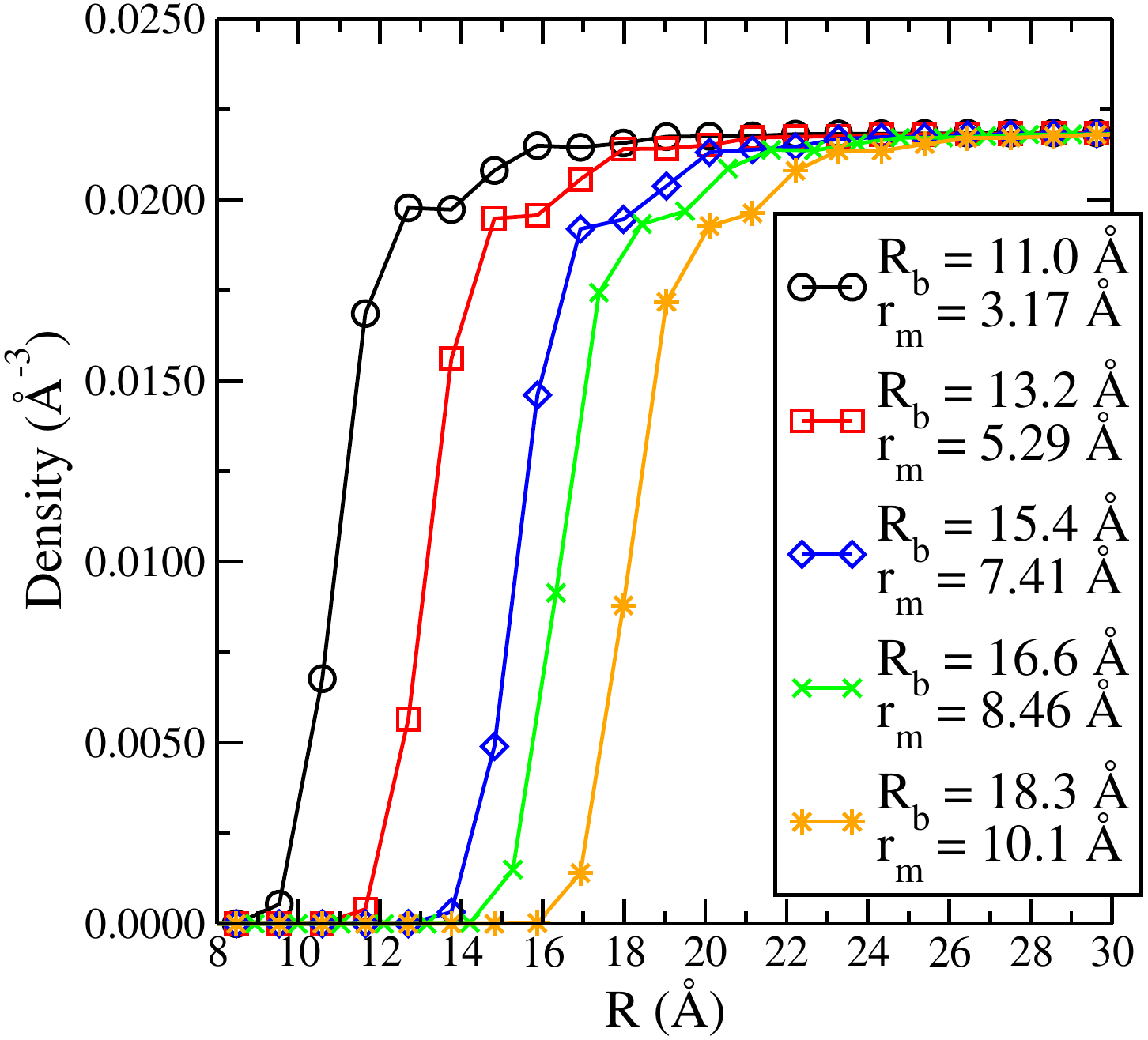}
\caption{Liquid density profiles at zero pressure 
for bubbles with varying diameter (determined by parameter $r_m$ in Eq. (\ref{eq0})). The bubble radii $R_b$ were calculated according to Eq. (\ref{eq5}).}
\label{fig3}
\end{figure}

The equilibrium liquid structures around the bubbles studied in this work are summarized in Fig. \ref{fig3}. Since the interface between the bubble and the liquid has a finite width, care must be taken to specify the bubble radius unambiguously. We compute the bubble radius, $R_b$, from the liquid profile by using the following balance equation for the bubble interface mass distribution:\cite{eloranta7}
\begin{equation}
\int_0^{R_b}\rho(r)d^3r = \int_{R_b}^\infty \left(\rho_0 - \rho(r)\right)d^3r
\label{eq5}
\end{equation}
where $R_b$ is called the mass barycenter of the interface. Note, however, that in the presence of bound solvent layers around the bubble (as it occurs, e.g., for positive ions in liquid $^4$He), this definition would become ill-defined. In such cases the hydrodynamic added mass given by Eq. (\ref{eq0c}) becomes a more meaningful measure of the object size (or mass).\cite{eloranta6} The $R_b$ values for the bubbles employed in this study are indicated in Fig. \ref{fig3}.

\begin{figure}
\includegraphics[scale=.6]{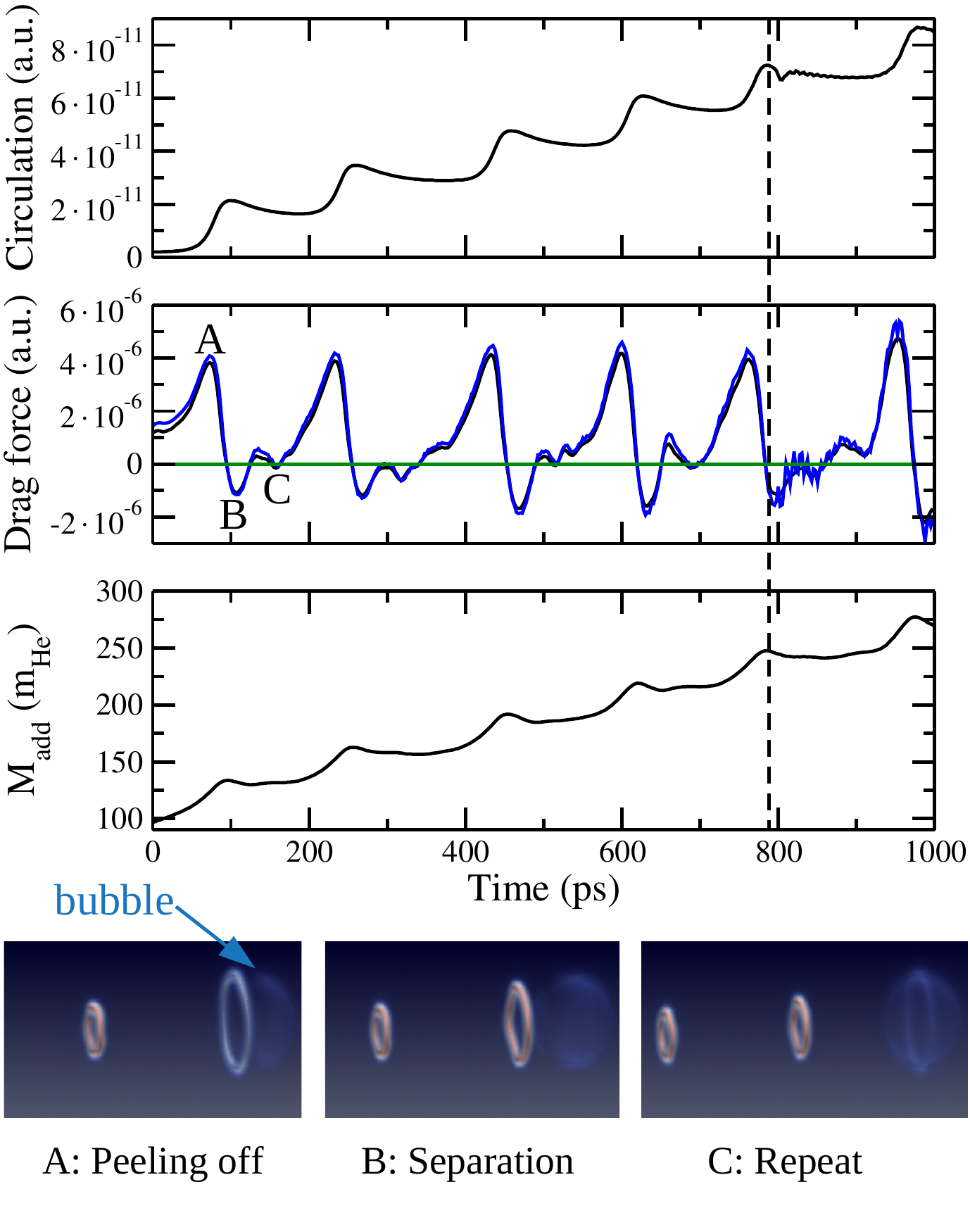}
\caption{Integrated liquid circulation corresponding to Eq. (\ref{eq-circ2}) (top graph), longitudinal and maximum transverse forces (green) on the bubble from Eq. (\ref{eq0a}) in black and from Eq.(\ref{eq0b}) in blue (middle graph), and hydrodynamic added mass obtained from Eq. (\ref{eq0c}) (bottom graph). The data shown corresponds to a bubble with $r_m$ = 3.17 \AA{}, velocity 75.4 m/s and zero external pressure. The dashed line indicates the point when the 
sound waves emitted after the warm-up period pass through the periodic boundaries and reach the bubble again,
thus introducing artifacts in the simulated quantities. 
The volume plots of Eq. (\ref{eq-circ}) at the bottom show the time evolution of vortex ring emission at specified points in time as identified by labels A, B, and C. The leftmost vortex ring in these volume plots is a left-over from the previous emission cycle.}
\label{fig4}
\end{figure}

The quantities from Eqs. (\ref{eq-circ2}), (\ref{eq0a}), (\ref{eq0b}), and (\ref{eq0c}) were recorded as a function of time at every 2.5 ps during each simulation after the warm-up period was completed. 
As an example, these data for a bubble corresponding to $r_m = 3.17$ \AA{} are shown in Fig. \ref{fig4}. The onset of vortex ring emission is visible in the total liquid circulation (periodic steps), the drag force (sudden increase in drag force followed by a drop off and a tailing negative impulse), and the step-wise increases in the hydrodynamic mass that tends to level off in the long-time limit. Although not clearly visible in Fig. \ref{fig4}, the small aperiodicity present in the oscillatory features is due to correlated multi-vortex ring emission events (e.g., two vortex rings emitted back to back). Farther behind the bubble, the correlated vortex rings tend to leapfrog each other as they move downstream. This is the three-dimensional equivalent of the rotating ``vortex dimers" observed previously in 2-D simulations.\cite{pi2,sasaki1} The main vortex ring emission steps are demonstrated in the bottom part of Fig. \ref{fig4}. After a vortex ring is peeled off of the bubble, it shrinks slightly, then fully separates from the bubble, and finally the vortex ring emission cycle repeats over.

As an indicator of stability of the calculation, the transverse drag force components remain negligibly small (less than \textit{ca.} $10^{-10}$ a.u.; green line in Fig. \ref{fig4}). Note that this symmetry would already break within the first 200 ps if single precision floating point numbers were used in the simulation. With the currently applied simulation box size ($512\times 256\times 256$), the emitted waves re-enter the bubble region at approximately 800 ps. This is clearly visible in the drag force and the circulation (dashed line in Fig. \ref{fig4}). The small irregularities appearing already before this point are likely caused by the employed time integration scheme and possibly also due to the accuracy of double precision floating point numbers. The two definitions of drag force (i.e., Eqs. (\ref{eq0a}) and (\ref{eq0b})) follow each other within the numerical accuracy as demonstrated in Fig. \ref{fig4}. Note that Eq. (\ref{eq0a}) is sensitive to the local surroundings of the bubble whereas in Eq. (\ref{eq0b}) the drag is determined by the time derivative of the (global) momentum transfer to the liquid.

\begin{figure}
\includegraphics[scale=.5]{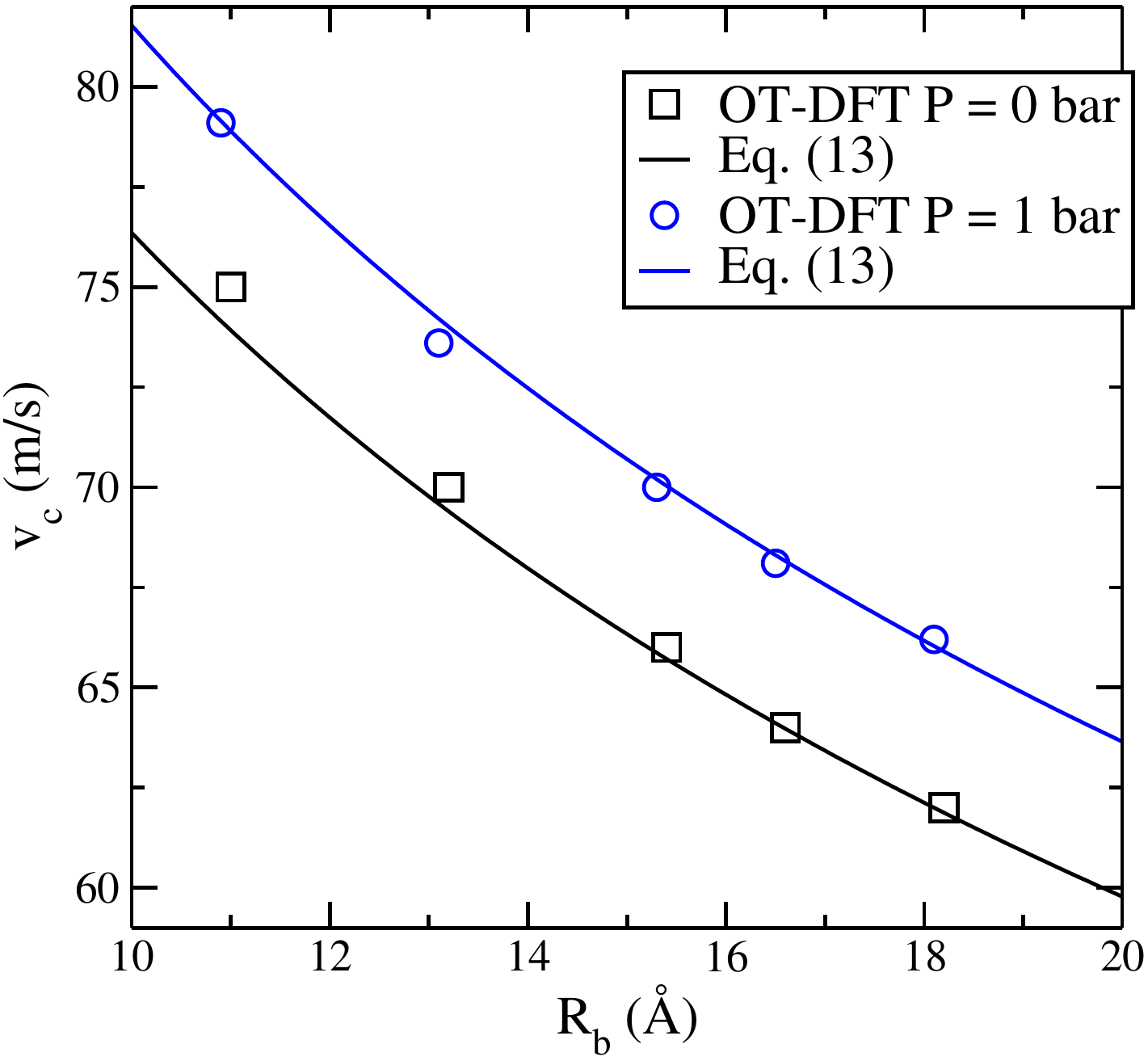}
\caption{Critical velocity ($v_c$) for vortex ring emission as a function of bubble radius $R_b$ at $P = 0$ bar (black squares) and $P = 1$ bar (blue circles). The continuous lines correspond to Eq. (\ref{eq-ratio}) with $a_0 = 0.74$ (black line; value from Ref. \onlinecite{eloranta2}) \AA{} and $a_0 = 0.67$ \AA{} (blue line; value from least squares fit). Note that the uncertainty in the OT-DFT calculated values of $v_c$ due to the periodic boundaries is $\pm 1$ m/s.}
\label{fig5}
\end{figure}

The onset of vortex ring emission appears always abruptly at a characteristic critical velocity value that is determined by the bubble radius and the applied external pressure. Variation of the critical velocity ($v_c$) as a function of the bubble radius at two different external pressures is shown in Fig. \ref{fig5}. The data displays the following important trends for the critical velocity: 1) it decreases as a function of the bubble radius and 2) it increases with increasing external pressure. For the largest bubble ($r_m = 10.1$ \AA{}), additional calculations were carried out using extended spatial grids ($512\times 512\times 512$ and $1024\times 256\times 256$) to determine the possible effect of the periodic boundaries on the system. The critical velocity difference between the standard and extended grids was approximately 1 m/s, which is within the accuracy of $v_c$ determination imposed by the periodic grid.  

After the critical velocity threshold has been exceeded, the vortex ring emission frequency increases rapidly with the flow velocity and the system quickly becomes unstable (i.e., breaking of the cylindrical symmetry). In this case, as opposed to the discrete steps seen in the graphs of Fig. \ref{fig4}, the data from such symmetry broken simulations do not show any systematic structure. Plotting the liquid circulation using Eq. (\ref{eq-circ}), reveals that the discrete vortex ring emission events are still taking place, but the rings no longer form symmetrically around the bubble (see Fig. \ref{fig6}). During this time, the total circulation increases approximately monotonically as function of time. Furthermore, complex behavior, such as vortex ring leapfrogging, breaking of vortex rings into smaller rings, and fusion of two vortex rings into larger ones, can also be observed at high velocities (cf. Fig. \ref{fig6}). The likely origin of the symmetry breaking (and transition to randomized turbulent-like flow) in our calculations is the presence of a small amount of ``numerical noise" that can be thought to play the role of randomized thermal motion.

\section{Discussion}

The Gross-Pitaevskii model, despite its shortcomings, is often applied to model superfluid helium.\cite{berloff3,busta,salman1} As GP is known to model only the phonon branch of superfluid helium dispersion relation, it is typically employed with parametrization that matches the speed of sound (GP Model 1 in Fig. \ref{fig2}). However, as previous studies have noted,\cite{salman1} this does not yield results that are in agreement with known ion solvation structures. Therefore, an alternative parametrization, which was inspired by attempting to match the available experimental data, has been used (GP Model 2 in Fig. \ref{fig2}). Although GP Model 2 no longer produces the correct speed of sound, it can model the vortex core structure with a reasonable accuracy (see Fig. \ref{fig1}). Since vortex ring energetics determines the onset of the emission process, GP Model 2 is expected to reproduce the related experimental quantities with much better accuracy than GP Model 1. However, other phenomena (e.g., dissipation through emission of sound, roton emission) cannot be described correctly by this model. The general features of vortex ring emission obtained from GP Model 2 appear very similar to those observed in the present OT-DFT simulations.

\begin{figure*}
\begin{center}
\includegraphics[scale=.38]{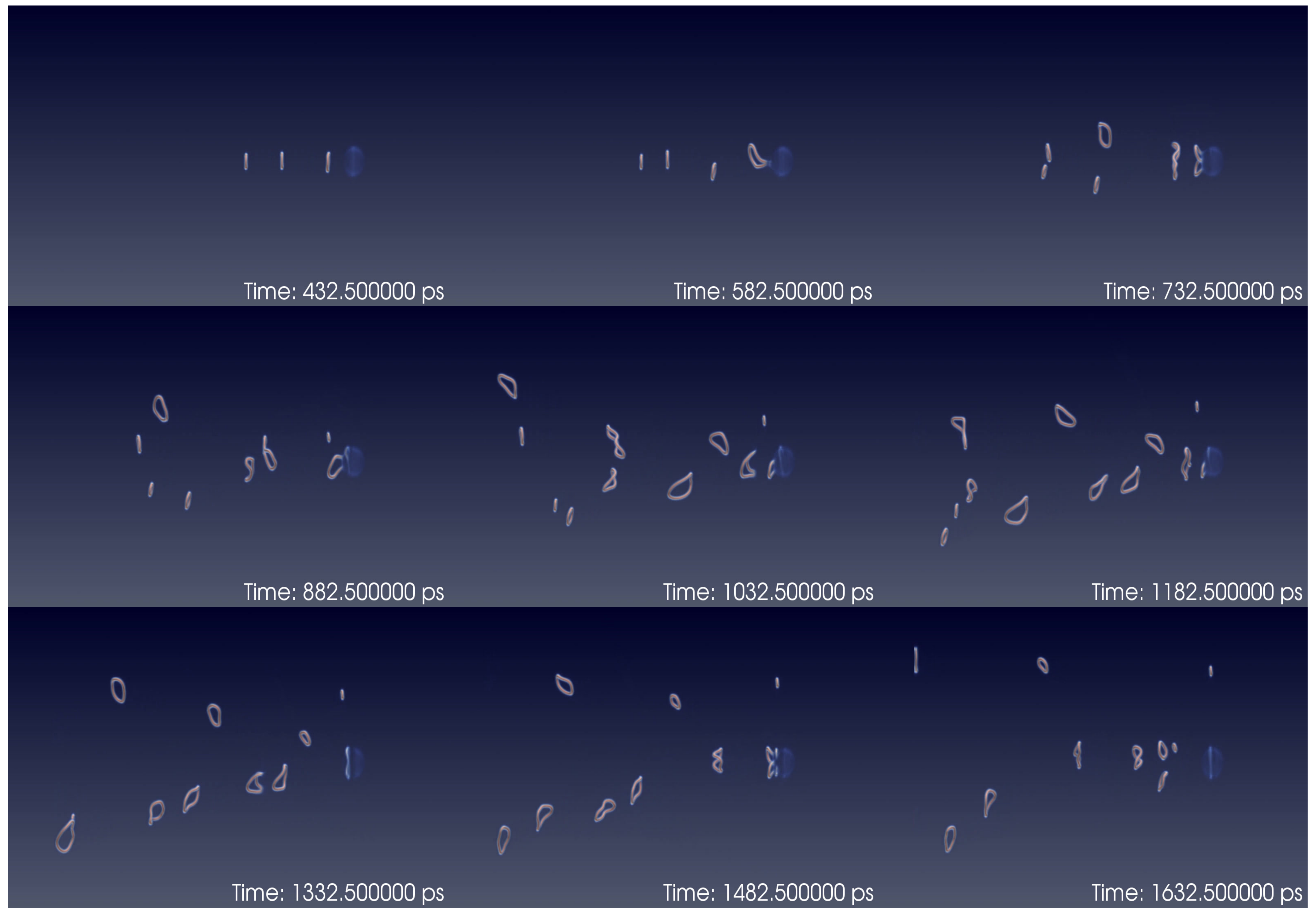}
\end{center}
\caption{Snapshots of liquid current circulation from Eq. (\ref{eq-circ}) at indicated times for a bubble with $r_m = 3.17$ \AA{} traveling at 77.3 m/s. Note that the flow velocity is higher than the critical velocity for vortex ring emission (75.4 m/s). The length scale in each frame is identical and set by the bubble with $R_b = 11.0$ \AA{}.}
\label{fig6}
\end{figure*}

One crucial difference between local GP-based and non-local DFT models (e.g., OT-DFT and the model described in Ref. \onlinecite{berloff1}) is that the latter can describe the roton branch of the dispersion relation (cf. Fig \ref{fig2}). Since the early work of Landau,\cite{wilks1} the onset of dissipation in superfluid helium was suggested to be due to the coupling of the moving object to rotons (i.e., roton emission). Accordingly, the Landau critical velocity ($v_L$) for dissipative motion was defined by the slope of the line that connects the origin and the roton minimum in the dispersion relation (approx. 59 m/s).\cite{wilks1} By coincidence, this value closely matches the critical velocity for vortex ring emission for the electron bubble at zero pressure.\cite{borghesani1,anci1,guo,salman1} However, much lower values (down to few cm/s) are typically seen at larger length scales,\cite{wilks1} which clearly shows that $v_L$ must scale with the object size, as was also shown by recent numerical simulations of a moving wire in superfluid $^4$He in 2-D geometry.\cite{pi2} Comparison of our OT-DFT calculations with the previous GP results  shows that the same vortex ring emission process can take place in both models, which clearly demonstrates that rotons do not play a major role in the onset of dissipation nor in the existence of the critical velocity. A similar note was also made by Balibar based on the lack of roton excitations in superfluid gases.\cite{balibar1} Furthermore, by increasing the external pressure, the value of the critical velocity appears to increase (cf. Fig. \ref{fig5}) whereas, at the same time, the roton minimum energy is lowered. This is also in contradiction with the Landau's criterion, according to which the critical velocity is expected to decrease with the decreasing roton gap. Note that the increase in the critical velocity with pressure, as shown in Fig. \ref{fig5}, has also been observed experimentally for the electron bubble.\cite{bowley1,salman1} Lastly, our simulations show that the critical velocity decreases as a function of the object size as demonstrated in Fig. \ref{fig5} (see also Fig. \ref{fig3}), which was also seen previously in 2-D He-DFT simulations. \cite{pi2} While the present results are in agreement with the above mentioned general experimental findings, they are clearly incompatible with the original concept of $v_L$. Therefore, we conclude that the observed critical velocity threshold is not related to roton emission and, in general, there does not seem to be any deeper connection between the two types of excitations apart from virtual rotons contributing to the liquid density oscillations around vortex lines (cf. Fig. \ref{fig1}). In fact, this general conclusion was already suggested by Roberts and Berloff in Ref. \onlinecite{roberts3}: ``\textit{Nevertheless it could, through an artificial example, provide strong indications that roton emission and vortex nucleation are different processes, the former being connected to the Landau critical velocity, and the latter to the speed of sound.}" However, the speed of sound in the OT-DFT model is \textit{ca.} 230 m/s and therefore the critical velocity for vortex ring emission does not appear to be related to this quantity either. Although the flow velocity is expected to be larger at the sides of the bubble, this difference is not sufficiently large that speed of sound could be reached around the bubble waist. For example, for incompressible inviscid flow, the liquid velocity at the sides is larger than the flow velocity by a factor of 3/2. Accounting for this difference would predict that a flow velocity of \textit{ca.} 150 m/s at the front would yield a waist velocity that reaches the speed of sound. However, this flow velocity value is much larger than the critical velocity thresholds observed in the OT-DFT simulations. Moreover, this model would predict that the critical velocity is {\it independent} of the bubble radius, at variance with our present results. Based on the convention outlined in Ref. \onlinecite{varoquaux1}, the velocity limit for dissipation in 
the present case corresponds to Feynman critical velocity (vortex emission) rather than to Landau critical velocity (roton emission).

Once the flow velocity reaches the Feynman critical limit, sudden nearly periodic variations in the drag force are observed. The underlying vortex ring emission events are characterized by oscillating drag force where the rising edge of the peak (cf. middle graph in Fig. \ref{fig4}) correlates with the formation of the vortex ring around the bubble (cf. bottom section of the same figure). Correspondingly, the decrease in the drag force peak is related to the detachment of the vortex ring that is finally accompanied by a small negative drag force due to the vortex-bubble separation and vortex ring shrinking processes. At the times of vortex ring shedding, both the total liquid circulation as well as the hydrodynamic added mass show step-wise increments. All these three quantities can be used to count the vortex shedding events during simulations, provided that the cylindrical symmetry is preserved (i.e., the steps can be identified). 

When the flow velocity is close to the critical velocity threshold, fully symmetric smooth vortex ring shedding takes place as shown in the volume plots of Fig. \ref{fig4}. However, by increasing the velocity by just a few m/s, the cylindrical symmetry is quickly lost and asymmetric vortex rings detach gradually from the bubble as demonstrated in Fig. \ref{fig6}. These two scenarios are known as the ``peeling" and ``girdling"  mechanisms, correspondingly.\cite{borghesani1,bowley1} In the latter case, based on the observed dynamic evolution of the vortex rings in the simulations, periodic excitations (e.g., Kelvin waves) around the rings were responsible for the initial geometric distortions. The flow velocity range, where the cylindrical symmetry was preserved (i.e., vortex peeling), was observed to be wider for small bubbles than for large bubbles. This observation may be related to the fact that the energy spectrum for the excited modes on vortex rings must be discrete due to the presence of the cyclic boundary condition. For larger bubbles the quantized energy levels must be closer to each other and therefore easier to excite, whereas the rings formed around smaller bubbles must overcome a larger energy gap. For circular Kelvin waves, the quantized energy level structure can be estimated based on the known dispersion relation (infinitely long vortex line) as well as the restriction imposed by the cyclic boundary condition:\cite{procaccia,pitaevskii}
\begin{equation}
\omega(k) \approx \frac{\Gamma k^2}{4\pi}\ln\left(ka_0\right)\textnormal{ with }k = \frac{n}{R_b}
\label{eq-kelvin}
\end{equation}
where circulation $\Gamma = h / m_{He} \approx 9.97\times 10^{-3}$ cm$^2$/s, vortex core parameter $a_0 \approx 10^{-8}$ cm, and $k$ is the wave number (cm$^{-1}$). The latter quantity is quantized due the cyclic boundary condition, which depends on the bubble radius and integer quantum number $n = 0, 1, 2, ...$. Note that the value of $\omega$ above is negative, which reflects the relative direction of liquid rotation.\cite{pitaevskii} It can be estimated from Eq. (\ref{eq-kelvin}) that the lowest excitation energy for a vortex ring formed around a bubble with $R_b = 10$ \AA{} is 1.4 K whereas for $R_b = 20$ \AA{} this is only 0.45 K. Thus it would be more difficult to excite the symmetry breaking modes in smaller bubbles vs. the larger ones. For objects of macroscopic dimensions, the excitation spectrum becomes essentially continuous and the system would therefore be subject to spontaneous symmetry breaking.

\begin{figure}
\includegraphics[scale=.8]{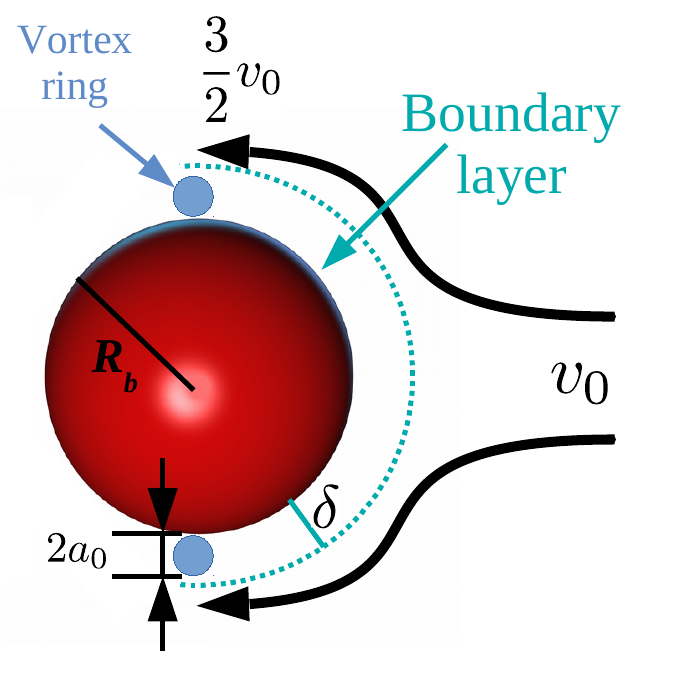}
\caption{Diagram of superfluid helium flow past a spherical object with radius $R_b$ at velocity $v_0$. The equatorial velocity for incompressible laminar flow is $\frac{3}{2}v_0$, the boundary layer thickness is denoted by $\delta$ (see text for explanation), and $a_0$ is the vortex core parameter.}
\label{fig7}
\end{figure}

To obtain a simple estimate for the onset of 
vortex ring emission, we consider the ratio between the kinetic energy present in the boundary layer around the bubble (see Fig. \ref{fig7}) and the energy required to create a single vortex ring with the same radius as the bubble. The boundary layer energy can be estimated by:
\begin{equation}
E_b(R, \vec{v}) = \frac{1}{2} M(R_b) \left|\vec{v}\right|^2 \approx \frac{3}{4} M(R_b) v_0^2 
\label{eq2}
\end{equation}
where $\vec{v}$ is the fluid velocity field. Note that the actual flow velocity is position dependent, but above we have arbitrarily chosen it to correspond to the equatorial velocity of incompressible laminar flow around a sphere: $\left|\vec{v}\right| = \frac{3}{2}v_0$ where $v_0$ is the flow velocity at the front. It turns out that this choice does not significantly affect the outcome of the model as it can be approximately incorporated into other model parameters. The helium mass of the boundary layer is given by:
\begin{equation}
M(R_b) = \frac{1}{2}\times 4\pi R_b^2\delta\rho_0
\label{eq3}
\end{equation}
with $R_b$ being the bubble radius, the empirical parameter $\delta = 5a_0$ describes the thickness of the boundary layer around the bubble, and $\rho_0$ is the bulk liquid density (\textit{ca.} 145 kg/m$^3$ at zero temperature and pressure). The factor 1/2 in front of Eq. (\ref{eq3}) considers only the front part of the bubble with respect to the flow as the backside does not contribute to vortex ring formation. Note that Eq. (\ref{eq3}) assumes that the interaction potential between the bubble and the liquid is repulsive such that there is no pronounced solvent layer structure around it (i.e., ``slipping" boundary condition). 
The energy required for creating a single vortex ring with radius $R_b$ is:\cite{donnelly2}
\begin{equation}
E_v(R_b) = \frac{2\pi\hbar^2}{m_{\textnormal{He}}^2}\rho_0 R_b\left[\ln\left(\frac{8R_b}{a_0}\right) - 1.615\right]
\label{eq4}
\end{equation}
The boundary layer kinetic energy, $E_b$, and the energy required for creating the vortex ring, $E_v$, should be equal at the critical velocity ($v_c$):
\begin{equation}
\frac{E_b}{E_v} = 1
\label{eq-ratio}
\end{equation}
This condition provides an implicit relationship between $v_c$ and the bubble radius $R_b$. Note that the bulk liquid density ($\rho_0$) cancels in this ratio and hence the pressure dependence of $v_c$ is solely determined by the variations in the vortex core parameter $a_0$, which should decrease when the external pressure increases. This behavior is in agreement with the OT-DFT data shown in Fig. \ref{fig5} as well as the experimental data described in the literature.\cite{bowley1,salman1} The relationship between the critical velocity and the sphere radius is non-linear, especially at small values of $R_b$. Based on the previous suggestion that $\textnormal{Re}_s\propto v R_b$ and that the vorticity should appear at some characteristic value (Re$_c$), the critical velocity should scale as $v_c \propto \textnormal{Re}_c / R_b$. However, this form does not fit the OT-DFT data shown in Fig. \ref{fig5}, but would instead require, e.g., a modified form: $v_c \propto \textnormal{Re}_c / v + C$ where $C$ is a constant. 

Finally, we note that the critical velocity predicted at the radius corresponding approximately to an electron bubble ($R \approx 18.5$ \AA{}), we obtain a value of \textit{ca.} 61 m/s, which is slightly larger than that observed experimentally (56 m/s)\cite{bowley1} and those predicted by earlier numerical simulations based on GP Model 2.\cite{salman1} This small difference is likely due to the missing KC and BF terms in the basic OT-DFT functional employed in this work. These terms only provide a minor contribution to vortex ring energetics and the density modulations around the vortex core. Note also that the ``rigid bubble" potential of Eq. (\ref{eq0}) partially neglects the possible non-spherical distortions of the bubble induced by the flow which, as discussed previously, may result in early appearance of vortex rings. Considering these minor approximations, the quantitative agreement with the experimental critical value is very good.

\section{Conclusions}

We have successfully modeled vortex ring emission by moving spherical heliophobic bubbles in superfluid $^4$He at 0 K by TD-DFT. Provided that the cylindrical symmetry
is preserved, liquid current circulation, drag force, and hydrodynamic mass were shown to be useful observables for characterizing the underlying liquid dynamics. At high vortex emission
rates, the cylindrical symmetry of the system tends to break due to excitation of circular Kelvin wave modes by the unavoidable numerical noise present in the calculations. The complex
vortex ring dynamics appearing downstream includes vortex ring fission and reconnection events that are thought to be the basic ingredients of quantum turbulence. Since the onset of
dissipation in the present case is due to the formation of quantized vortex rings rather than roton emission, this threshold should be called Feynman critical velocity rather
than Landau critical velocity. Our calculations further show that the Feynman threshold is not directly related to the liquid speed of sound, but can rather be
rationalized by considering the ratio between the boundary layer kinetic and vortex ring formation energies. 
According to this model, the increase in critical velocity as a function of
pressure is caused by compression of the vortex core.
Although the current calculations only employ an artificial repulsive potential, parameter values that reproduce the correct electron bubble
geometry result in a similar critical velocity as observed experimentally.


\begin{thebibliography}{58}
\expandafter\ifx\csname natexlab\endcsname\relax\def\natexlab#1{#1}\fi
\expandafter\ifx\csname bibnamefont\endcsname\relax
  \def\bibnamefont#1{#1}\fi
\expandafter\ifx\csname bibfnamefont\endcsname\relax
  \def\bibfnamefont#1{#1}\fi
\expandafter\ifx\csname citenamefont\endcsname\relax
  \def\citenamefont#1{#1}\fi
\expandafter\ifx\csname url\endcsname\relax
  \def\url#1{\texttt{#1}}\fi
\expandafter\ifx\csname urlprefix\endcsname\relax\def\urlprefix{URL }\fi
\providecommand{\bibinfo}[2]{#2}
\providecommand{\eprint}[2][]{\url{#2}}

\bibitem[{\citenamefont{Borghesani}(2007)}]{borghesani1}
\bibinfo{author}{\bibfnamefont{A.~F.} \bibnamefont{Borghesani}},
  \emph{\bibinfo{title}{Ions and electrons in liquid helium}}
  (\bibinfo{publisher}{Oxford Science Publications}, \bibinfo{address}{Oxford},
  \bibinfo{year}{2007}).

\bibitem[{\citenamefont{Williams}(1957)}]{williams1}
\bibinfo{author}{\bibfnamefont{R.~L.} \bibnamefont{Williams}},
  \bibinfo{journal}{Can. J. Phys.} \textbf{\bibinfo{volume}{35}},
  \bibinfo{pages}{134} (\bibinfo{year}{1957}).

\bibitem[{\citenamefont{Careri et~al.}(1959)\citenamefont{Careri, Scaramuzzi,
  and Thomson}}]{thomson1}
\bibinfo{author}{\bibfnamefont{G.}~\bibnamefont{Careri}},
  \bibinfo{author}{\bibfnamefont{F.}~\bibnamefont{Scaramuzzi}},
  \bibnamefont{and} \bibinfo{author}{\bibfnamefont{J.~O.}
  \bibnamefont{Thomson}}, \bibinfo{journal}{Nuovo Cimento}
  \textbf{\bibinfo{volume}{13}}, \bibinfo{pages}{186} (\bibinfo{year}{1959}).

\bibitem[{\citenamefont{Reif and Meyer}(1960)}]{meyer1}
\bibinfo{author}{\bibfnamefont{F.}~\bibnamefont{Reif}} \bibnamefont{and}
  \bibinfo{author}{\bibfnamefont{L.}~\bibnamefont{Meyer}},
  \bibinfo{journal}{Phys. Rev.} \textbf{\bibinfo{volume}{119}},
  \bibinfo{pages}{1164} (\bibinfo{year}{1960}).

\bibitem[{\citenamefont{Muirhead et~al.}(1984)\citenamefont{Muirhead, Vinen,
  and Donnelly}}]{muirhead1}
\bibinfo{author}{\bibfnamefont{C.}~\bibnamefont{Muirhead}},
  \bibinfo{author}{\bibfnamefont{W.}~\bibnamefont{Vinen}}, \bibnamefont{and}
  \bibinfo{author}{\bibfnamefont{R.}~\bibnamefont{Donnelly}},
  \bibinfo{journal}{Phil. Trans. Roy. Soc. A} \textbf{\bibinfo{volume}{311}},
  \bibinfo{pages}{433} (\bibinfo{year}{1984}).

\bibitem[{\citenamefont{Ancilotto
  et~al.}(2017{\natexlab{a}})\citenamefont{Ancilotto, Barranco, Coppens,
  Eloranta, Halberstadt, Hernando, Mateo, and Pi}}]{dftreview}
\bibinfo{author}{\bibfnamefont{F.}~\bibnamefont{Ancilotto}},
  \bibinfo{author}{\bibfnamefont{M.}~\bibnamefont{Barranco}},
  \bibinfo{author}{\bibfnamefont{F.}~\bibnamefont{Coppens}},
  \bibinfo{author}{\bibfnamefont{J.}~\bibnamefont{Eloranta}},
  \bibinfo{author}{\bibfnamefont{N.}~\bibnamefont{Halberstadt}},
  \bibinfo{author}{\bibfnamefont{A.}~\bibnamefont{Hernando}},
  \bibinfo{author}{\bibfnamefont{D.}~\bibnamefont{Mateo}}, \bibnamefont{and}
  \bibinfo{author}{\bibfnamefont{M.}~\bibnamefont{Pi}}, \bibinfo{journal}{Int.
  Rev. Phys. Chem.} \textbf{\bibinfo{volume}{36}}, \bibinfo{pages}{621}
  (\bibinfo{year}{2017}{\natexlab{a}}).

\bibitem[{\citenamefont{Aitken et~al.}(2016)\citenamefont{Aitken, Bonifaci, von
  Haeften, and Eloranta}}]{eloranta1}
\bibinfo{author}{\bibfnamefont{F.}~\bibnamefont{Aitken}},
  \bibinfo{author}{\bibfnamefont{N.}~\bibnamefont{Bonifaci}},
  \bibinfo{author}{\bibfnamefont{K.}~\bibnamefont{von Haeften}},
  \bibnamefont{and} \bibinfo{author}{\bibfnamefont{J.}~\bibnamefont{Eloranta}},
  \bibinfo{journal}{J. Chem. Phys.} \textbf{\bibinfo{volume}{145}},
  \bibinfo{pages}{044105} (\bibinfo{year}{2016}).

\bibitem[{\citenamefont{Wilks}(1967)}]{wilks1}
\bibinfo{author}{\bibfnamefont{J.}~\bibnamefont{Wilks}},
  \emph{\bibinfo{title}{The Properties of Liquid and Solid Helium}}
  (\bibinfo{publisher}{Clarendon Press}, \bibinfo{address}{Oxford},
  \bibinfo{year}{1967}).

\bibitem[{\citenamefont{Barenghi et~al.}(2001)\citenamefont{Barenghi, Donnelly,
  and Vinen}}]{vinen1}
\bibinfo{author}{\bibfnamefont{C.~F.} \bibnamefont{Barenghi}},
  \bibinfo{author}{\bibfnamefont{R.~J.} \bibnamefont{Donnelly}},
  \bibnamefont{and} \bibinfo{author}{\bibfnamefont{W.~F.} \bibnamefont{Vinen}},
  \emph{\bibinfo{title}{Quantized Vortex Dynamics and Superfluid Turbulence}}
  (\bibinfo{publisher}{Springer}, \bibinfo{address}{New York},
  \bibinfo{year}{2001}).

\bibitem[{\citenamefont{Villois and Salman}(2018)}]{salman1}
\bibinfo{author}{\bibfnamefont{A.}~\bibnamefont{Villois}} \bibnamefont{and}
  \bibinfo{author}{\bibfnamefont{H.}~\bibnamefont{Salman}},
  \bibinfo{journal}{Phys. Rev. B} \textbf{\bibinfo{volume}{97}},
  \bibinfo{pages}{094507} (\bibinfo{year}{2018}).

\bibitem[{\citenamefont{Berloff and Roberts}(2001{\natexlab{a}})}]{roberts1}
\bibinfo{author}{\bibfnamefont{N.~G.} \bibnamefont{Berloff}} \bibnamefont{and}
  \bibinfo{author}{\bibfnamefont{P.~H.} \bibnamefont{Roberts}},
  \bibinfo{journal}{J. Phys. A} \textbf{\bibinfo{volume}{34}},
  \bibinfo{pages}{81} (\bibinfo{year}{2001}{\natexlab{a}}).

\bibitem[{\citenamefont{Frisch et~al.}(1992)\citenamefont{Frisch, Pomeau, and
  Rica}}]{rica1}
\bibinfo{author}{\bibfnamefont{T.}~\bibnamefont{Frisch}},
  \bibinfo{author}{\bibfnamefont{Y.}~\bibnamefont{Pomeau}}, \bibnamefont{and}
  \bibinfo{author}{\bibfnamefont{S.}~\bibnamefont{Rica}},
  \bibinfo{journal}{Phys. Rev. Lett.} \textbf{\bibinfo{volume}{69}},
  \bibinfo{pages}{1644} (\bibinfo{year}{1992}).

\bibitem[{\citenamefont{Navon et~al.}(2016)\citenamefont{Navon, Gaunt, Smith,
  and Hadzibabic}}]{hadzibabic1}
\bibinfo{author}{\bibfnamefont{N.}~\bibnamefont{Navon}},
  \bibinfo{author}{\bibfnamefont{A.~L.} \bibnamefont{Gaunt}},
  \bibinfo{author}{\bibfnamefont{R.~P.} \bibnamefont{Smith}}, \bibnamefont{and}
  \bibinfo{author}{\bibfnamefont{Z.}~\bibnamefont{Hadzibabic}},
  \bibinfo{journal}{Nature} \textbf{\bibinfo{volume}{539}}, \bibinfo{pages}{72}
  (\bibinfo{year}{2016}).

\bibitem[{\citenamefont{Berloff and Svistunov}(2002)}]{svistunov1}
\bibinfo{author}{\bibfnamefont{N.~G.} \bibnamefont{Berloff}} \bibnamefont{and}
  \bibinfo{author}{\bibfnamefont{B.~V.} \bibnamefont{Svistunov}},
  \bibinfo{journal}{Phys. Rev. A} \textbf{\bibinfo{volume}{66}},
  \bibinfo{pages}{013603} (\bibinfo{year}{2002}).

\bibitem[{\citenamefont{Kobayashi and Tsubota}(2005)}]{tsubota1}
\bibinfo{author}{\bibfnamefont{M.}~\bibnamefont{Kobayashi}} \bibnamefont{and}
  \bibinfo{author}{\bibfnamefont{M.}~\bibnamefont{Tsubota}},
  \bibinfo{journal}{Phys. Rev. Lett.} \textbf{\bibinfo{volume}{94}},
  \bibinfo{pages}{065302} (\bibinfo{year}{2005}).

\bibitem[{\citenamefont{Proment et~al.}(2009)\citenamefont{Proment, Nazarenko,
  and Onorato}}]{onorato1}
\bibinfo{author}{\bibfnamefont{D.}~\bibnamefont{Proment}},
  \bibinfo{author}{\bibfnamefont{S.}~\bibnamefont{Nazarenko}},
  \bibnamefont{and} \bibinfo{author}{\bibfnamefont{M.}~\bibnamefont{Onorato}},
  \bibinfo{journal}{Phys. Rev. A} \textbf{\bibinfo{volume}{80}},
  \bibinfo{pages}{051603} (\bibinfo{year}{2009}).

\bibitem[{\citenamefont{Mateo et~al.}(2015)\citenamefont{Mateo, Eloranta, and
  Williams}}]{eloranta2}
\bibinfo{author}{\bibfnamefont{D.}~\bibnamefont{Mateo}},
  \bibinfo{author}{\bibfnamefont{J.}~\bibnamefont{Eloranta}}, \bibnamefont{and}
  \bibinfo{author}{\bibfnamefont{G.~A.} \bibnamefont{Williams}},
  \bibinfo{journal}{J. Chem. Phys.} \textbf{\bibinfo{volume}{142}},
  \bibinfo{pages}{064510} (\bibinfo{year}{2015}).

\bibitem[{\citenamefont{Landau and Lifshitz}(2009)}]{lifshitz1}
\bibinfo{author}{\bibfnamefont{L.~D.} \bibnamefont{Landau}} \bibnamefont{and}
  \bibinfo{author}{\bibfnamefont{E.~M.} \bibnamefont{Lifshitz}},
  \emph{\bibinfo{title}{Fluid Mechanics (2nd ed.)}}
  (\bibinfo{publisher}{Butterworth and Heineman}, \bibinfo{address}{New York},
  \bibinfo{year}{2009}).

\bibitem[{\citenamefont{Feynman}(1955)}]{feynman1}
\bibinfo{author}{\bibfnamefont{R.~P.} \bibnamefont{Feynman}},
  \emph{\bibinfo{title}{Progr. Low Temp. Phys. (Vol. 1)}}
  (\bibinfo{publisher}{Elsevier}, \bibinfo{address}{New York},
  \bibinfo{year}{1955}).

\bibitem[{\citenamefont{Varoquaux}(2006)}]{varoquaux1}
\bibinfo{author}{\bibfnamefont{E.}~\bibnamefont{Varoquaux}},
  \bibinfo{journal}{C. R. Physique} \textbf{\bibinfo{volume}{7}},
  \bibinfo{pages}{1101} (\bibinfo{year}{2006}).

\bibitem[{\citenamefont{Nancolas et~al.}(1985)\citenamefont{Nancolas, Ellis,
  McClintock, and Bowley}}]{bowley1}
\bibinfo{author}{\bibfnamefont{G.~G.} \bibnamefont{Nancolas}},
  \bibinfo{author}{\bibfnamefont{T.}~\bibnamefont{Ellis}},
  \bibinfo{author}{\bibfnamefont{P.~V.~E.} \bibnamefont{McClintock}},
  \bibnamefont{and} \bibinfo{author}{\bibfnamefont{R.~M.}
  \bibnamefont{Bowley}}, \bibinfo{journal}{Nature}
  \textbf{\bibinfo{volume}{316}}, \bibinfo{pages}{797} (\bibinfo{year}{1985}).

\bibitem[{\citenamefont{Khalatnikov}(2000)}]{khalatnikov1}
\bibinfo{author}{\bibfnamefont{I.~M.} \bibnamefont{Khalatnikov}},
  \emph{\bibinfo{title}{An introduction to the theory of superfluidity}}
  (\bibinfo{publisher}{Westview Press}, \bibinfo{address}{Oxford},
  \bibinfo{year}{2000}).

\bibitem[{\citenamefont{Donnelly}(October, 2009)}]{donnelly1}
\bibinfo{author}{\bibfnamefont{R.~J.} \bibnamefont{Donnelly}},
  \bibinfo{journal}{Physics Today} p.~\bibinfo{pages}{34}
  (\bibinfo{year}{October, 2009}).

\bibitem[{\citenamefont{Bradley and Anderson}(2012)}]{anderson1}
\bibinfo{author}{\bibfnamefont{A.~S.} \bibnamefont{Bradley}} \bibnamefont{and}
  \bibinfo{author}{\bibfnamefont{B.~P.} \bibnamefont{Anderson}},
  \bibinfo{journal}{Phys. Rev. X} \textbf{\bibinfo{volume}{2}},
  \bibinfo{pages}{041001} (\bibinfo{year}{2012}).

\bibitem[{\citenamefont{Schoepe}(2018)}]{schoepe1}
\bibinfo{author}{\bibfnamefont{W.}~\bibnamefont{Schoepe}},
  \bibinfo{journal}{arXiv} \textbf{\bibinfo{volume}{1801.00593}}
  (\bibinfo{year}{2018}).

\bibitem[{\citenamefont{Reeves et~al.}(2015)\citenamefont{Reeves, Billam,
  Anderson, and Bradley}}]{reeves}
\bibinfo{author}{\bibfnamefont{M.~T.} \bibnamefont{Reeves}},
  \bibinfo{author}{\bibfnamefont{T.~P.} \bibnamefont{Billam}},
  \bibinfo{author}{\bibfnamefont{B.~P.} \bibnamefont{Anderson}},
  \bibnamefont{and} \bibinfo{author}{\bibfnamefont{A.~S.}
  \bibnamefont{Bradley}}, \bibinfo{journal}{Phys. Rev. Lett.}
  \textbf{\bibinfo{volume}{114}}, \bibinfo{pages}{155302}
  (\bibinfo{year}{2015}).

\bibitem[{\citenamefont{Schoepe}(2015)}]{schoepe2}
\bibinfo{author}{\bibfnamefont{W.}~\bibnamefont{Schoepe}},
  \bibinfo{journal}{JETP Lett.} \textbf{\bibinfo{volume}{102}},
  \bibinfo{pages}{105} (\bibinfo{year}{2015}).

\bibitem[{\citenamefont{Niemetz et~al.}(2017)\citenamefont{Niemetz, Hanninen,
  and Schoepe}}]{niemetz}
\bibinfo{author}{\bibfnamefont{M.}~\bibnamefont{Niemetz}},
  \bibinfo{author}{\bibfnamefont{R.}~\bibnamefont{Hanninen}}, \bibnamefont{and}
  \bibinfo{author}{\bibfnamefont{W.}~\bibnamefont{Schoepe}},
  \bibinfo{journal}{J. Low Temp. Phys.} \textbf{\bibinfo{volume}{187}},
  \bibinfo{pages}{195} (\bibinfo{year}{2017}).

\bibitem[{\citenamefont{Barenghi et~al.}(2014)\citenamefont{Barenghi, Skrbek,
  and Sreenivasan}}]{sreenivasan1}
\bibinfo{author}{\bibfnamefont{C.~F.} \bibnamefont{Barenghi}},
  \bibinfo{author}{\bibfnamefont{L.}~\bibnamefont{Skrbek}}, \bibnamefont{and}
  \bibinfo{author}{\bibfnamefont{K.~R.} \bibnamefont{Sreenivasan}},
  \bibinfo{journal}{PNAS} \textbf{\bibinfo{volume}{111}}, \bibinfo{pages}{4647}
  (\bibinfo{year}{2014}).

\bibitem[{\citenamefont{Dalfovo et~al.}(1995)\citenamefont{Dalfovo, Lastri,
  Pricaupenko, Stringari, and Treiner}}]{dalfovo}
\bibinfo{author}{\bibfnamefont{F.}~\bibnamefont{Dalfovo}},
  \bibinfo{author}{\bibfnamefont{A.}~\bibnamefont{Lastri}},
  \bibinfo{author}{\bibfnamefont{L.}~\bibnamefont{Pricaupenko}},
  \bibinfo{author}{\bibfnamefont{S.}~\bibnamefont{Stringari}},
  \bibnamefont{and} \bibinfo{author}{\bibfnamefont{J.}~\bibnamefont{Treiner}},
  \bibinfo{journal}{Phys. Rev. B} \textbf{\bibinfo{volume}{52}},
  \bibinfo{pages}{1193} (\bibinfo{year}{1995}).

\bibitem[{\citenamefont{Lehtovaara et~al.}(2004)\citenamefont{Lehtovaara,
  Kiljunen, and Eloranta}}]{eloranta3}
\bibinfo{author}{\bibfnamefont{L.}~\bibnamefont{Lehtovaara}},
  \bibinfo{author}{\bibfnamefont{T.}~\bibnamefont{Kiljunen}}, \bibnamefont{and}
  \bibinfo{author}{\bibfnamefont{J.}~\bibnamefont{Eloranta}},
  \bibinfo{journal}{J. Comp. Phys.} \textbf{\bibinfo{volume}{194}},
  \bibinfo{pages}{78} (\bibinfo{year}{2004}).

\bibitem[{\citenamefont{Berloff and Roberts}(2000{\natexlab{a}})}]{roberts2}
\bibinfo{author}{\bibfnamefont{N.~G.} \bibnamefont{Berloff}} \bibnamefont{and}
  \bibinfo{author}{\bibfnamefont{P.~H.} \bibnamefont{Roberts}},
  \bibinfo{journal}{Phys. Rev. B} \textbf{\bibinfo{volume}{63}},
  \bibinfo{pages}{024510} (\bibinfo{year}{2000}{\natexlab{a}}).

\bibitem[{\citenamefont{Amelio et~al.}(2018)\citenamefont{Amelio, Galli, and
  Reatto}}]{reatto}
\bibinfo{author}{\bibfnamefont{I.}~\bibnamefont{Amelio}},
  \bibinfo{author}{\bibfnamefont{D.~E.} \bibnamefont{Galli}}, \bibnamefont{and}
  \bibinfo{author}{\bibfnamefont{L.}~\bibnamefont{Reatto}},
  \bibinfo{journal}{arXiv} \textbf{\bibinfo{volume}{1802.07071}}
  (\bibinfo{year}{2018}).

\bibitem[{\citenamefont{Lehtovaara
  et~al.}(2018{\natexlab{a}})\citenamefont{Lehtovaara, Mateo, and
  Eloranta}}]{libdft}
\bibinfo{author}{\bibfnamefont{L.}~\bibnamefont{Lehtovaara}},
  \bibinfo{author}{\bibfnamefont{D.}~\bibnamefont{Mateo}}, \bibnamefont{and}
  \bibinfo{author}{\bibfnamefont{J.}~\bibnamefont{Eloranta}},
  \emph{\bibinfo{title}{{O}rsay-{T}rento density functional for superfluid
  helium}}, \bibinfo{howpublished}{Available on:
  \url{https://github.com/jmeloranta/libdft/}}
  (\bibinfo{year}{2018}{\natexlab{a}}).

\bibitem[{\citenamefont{Lehtovaara
  et~al.}(2018{\natexlab{b}})\citenamefont{Lehtovaara, Mateo, and
  Eloranta}}]{libgrid}
\bibinfo{author}{\bibfnamefont{L.}~\bibnamefont{Lehtovaara}},
  \bibinfo{author}{\bibfnamefont{D.}~\bibnamefont{Mateo}}, \bibnamefont{and}
  \bibinfo{author}{\bibfnamefont{J.}~\bibnamefont{Eloranta}},
  \emph{\bibinfo{title}{Library for 1-{D}, 2-{D} and 3-{D} grids}},
  \bibinfo{howpublished}{Available on:
  \url{https://github.com/jmeloranta/libgrid/}}
  (\bibinfo{year}{2018}{\natexlab{b}}).

\bibitem[{\citenamefont{Frigo and Johnson}(2005)}]{fftw}
\bibinfo{author}{\bibfnamefont{M.}~\bibnamefont{Frigo}} \bibnamefont{and}
  \bibinfo{author}{\bibfnamefont{S.}~\bibnamefont{Johnson}},
  \bibinfo{journal}{Proceedings of the IEEE} \textbf{\bibinfo{volume}{93}},
  \bibinfo{pages}{216} (\bibinfo{year}{2005}), \bibinfo{note}{special issue on
  ``Program Generation, Optimization, and Platform Adaptation''}.

\bibitem[{\citenamefont{Anon.}(2018)}]{cufft}
\bibinfo{author}{\bibnamefont{Anon.}}, \emph{\bibinfo{title}{CUFFT Library
  User's Guide}} (\bibinfo{publisher}{NVIDIA Corporation},
  \bibinfo{address}{Santa Clara, CA}, \bibinfo{year}{2018}).

\bibitem[{\citenamefont{Lehtovaara et~al.}(2007)\citenamefont{Lehtovaara,
  Toivanen, and Eloranta}}]{eloranta4}
\bibinfo{author}{\bibfnamefont{L.}~\bibnamefont{Lehtovaara}},
  \bibinfo{author}{\bibfnamefont{J.}~\bibnamefont{Toivanen}}, \bibnamefont{and}
  \bibinfo{author}{\bibfnamefont{J.}~\bibnamefont{Eloranta}},
  \bibinfo{journal}{J. Comp. Phys.} \textbf{\bibinfo{volume}{221}},
  \bibinfo{pages}{148} (\bibinfo{year}{2007}).

\bibitem[{\citenamefont{Press et~al.}(1999)\citenamefont{Press, Vetterling,
  Teukalosky, and Flannery}}]{numrep}
\bibinfo{author}{\bibfnamefont{W.}~\bibnamefont{Press}},
  \bibinfo{author}{\bibfnamefont{S.}~\bibnamefont{Vetterling}},
  \bibinfo{author}{\bibfnamefont{S.}~\bibnamefont{Teukalosky}},
  \bibnamefont{and} \bibinfo{author}{\bibfnamefont{B.}~\bibnamefont{Flannery}},
  \emph{\bibinfo{title}{Numerical Recipes in C (2nd ed.)}}
  (\bibinfo{publisher}{Cambridge University Press}, \bibinfo{address}{New
  York}, \bibinfo{year}{1999}).

\bibitem[{\citenamefont{Mateo et~al.}(2011)\citenamefont{Mateo, Jin, Pi, and
  Barranco}}]{mateo1}
\bibinfo{author}{\bibfnamefont{D.}~\bibnamefont{Mateo}},
  \bibinfo{author}{\bibfnamefont{D.}~\bibnamefont{Jin}},
  \bibinfo{author}{\bibfnamefont{M.}~\bibnamefont{Pi}}, \bibnamefont{and}
  \bibinfo{author}{\bibfnamefont{M.}~\bibnamefont{Barranco}},
  \bibinfo{journal}{J. Chem. Phys.} \textbf{\bibinfo{volume}{134}},
  \bibinfo{pages}{044507} (\bibinfo{year}{2011}).

\bibitem[{\citenamefont{Grau et~al.}(2006)\citenamefont{Grau, Barranco, Mayol,
  and Pi}}]{pi1}
\bibinfo{author}{\bibfnamefont{V.}~\bibnamefont{Grau}},
  \bibinfo{author}{\bibfnamefont{M.}~\bibnamefont{Barranco}},
  \bibinfo{author}{\bibfnamefont{R.}~\bibnamefont{Mayol}}, \bibnamefont{and}
  \bibinfo{author}{\bibfnamefont{M.}~\bibnamefont{Pi}}, \bibinfo{journal}{Phys.
  Rev. B} \textbf{\bibinfo{volume}{73}}, \bibinfo{pages}{064502}
  (\bibinfo{year}{2006}).

\bibitem[{\citenamefont{Eloranta and Apkarian}(2002)}]{eloranta5}
\bibinfo{author}{\bibfnamefont{J.}~\bibnamefont{Eloranta}} \bibnamefont{and}
  \bibinfo{author}{\bibfnamefont{V.~A.} \bibnamefont{Apkarian}},
  \bibinfo{journal}{J. Chem. Phys.} \textbf{\bibinfo{volume}{117}},
  \bibinfo{pages}{10139} (\bibinfo{year}{2002}).

\bibitem[{\citenamefont{Ancilotto et~al.}(2010)\citenamefont{Ancilotto,
  Barranco, and Pi}}]{anci1}
\bibinfo{author}{\bibfnamefont{F.}~\bibnamefont{Ancilotto}},
  \bibinfo{author}{\bibfnamefont{M.}~\bibnamefont{Barranco}}, \bibnamefont{and}
  \bibinfo{author}{\bibfnamefont{M.}~\bibnamefont{Pi}}, \bibinfo{journal}{Phys.
  Rev. B} \textbf{\bibinfo{volume}{82}}, \bibinfo{pages}{014517}
  (\bibinfo{year}{2010}).

\bibitem[{\citenamefont{Jin and Guo}(2010)}]{guo}
\bibinfo{author}{\bibfnamefont{D.}~\bibnamefont{Jin}} \bibnamefont{and}
  \bibinfo{author}{\bibfnamefont{W.}~\bibnamefont{Guo}},
  \bibinfo{journal}{Phys. Rev. B} \textbf{\bibinfo{volume}{82}},
  \bibinfo{pages}{094524} (\bibinfo{year}{2010}).

\bibitem[{\citenamefont{Ahrens et~al.}(2005)\citenamefont{Ahrens, Geveci, and
  Law}}]{paraview}
\bibinfo{author}{\bibfnamefont{J.}~\bibnamefont{Ahrens}},
  \bibinfo{author}{\bibfnamefont{B.}~\bibnamefont{Geveci}}, \bibnamefont{and}
  \bibinfo{author}{\bibfnamefont{C.}~\bibnamefont{Law}},
  \emph{\bibinfo{title}{Visualization Handbook}}
  (\bibinfo{publisher}{Elsevier}, \bibinfo{address}{Burlington, MA},
  \bibinfo{year}{2005}).

\bibitem[{\citenamefont{Ancilotto
  et~al.}(2017{\natexlab{b}})\citenamefont{Ancilotto, Barranco, Eloranta, and
  Pi}}]{pi2}
\bibinfo{author}{\bibfnamefont{F.}~\bibnamefont{Ancilotto}},
  \bibinfo{author}{\bibfnamefont{M.}~\bibnamefont{Barranco}},
  \bibinfo{author}{\bibfnamefont{J.}~\bibnamefont{Eloranta}}, \bibnamefont{and}
  \bibinfo{author}{\bibfnamefont{M.}~\bibnamefont{Pi}}, \bibinfo{journal}{Phys.
  Rev. B} \textbf{\bibinfo{volume}{96}}, \bibinfo{pages}{064503}
  (\bibinfo{year}{2017}{\natexlab{b}}).

\bibitem[{\citenamefont{Mateo and Eloranta}(2014)}]{eloranta6}
\bibinfo{author}{\bibfnamefont{D.}~\bibnamefont{Mateo}} \bibnamefont{and}
  \bibinfo{author}{\bibfnamefont{J.}~\bibnamefont{Eloranta}},
  \bibinfo{journal}{J. Phys. Chem. A} \textbf{\bibinfo{volume}{118}},
  \bibinfo{pages}{6407} (\bibinfo{year}{2014}).

\bibitem[{\citenamefont{Eloranta et~al.}(2002)\citenamefont{Eloranta,
  Schwentner, and Apkarian}}]{eloranta7}
\bibinfo{author}{\bibfnamefont{J.}~\bibnamefont{Eloranta}},
  \bibinfo{author}{\bibfnamefont{N.}~\bibnamefont{Schwentner}},
  \bibnamefont{and} \bibinfo{author}{\bibfnamefont{V.~A.}
  \bibnamefont{Apkarian}}, \bibinfo{journal}{J. Chem. Phys.}
  \textbf{\bibinfo{volume}{116}}, \bibinfo{pages}{4039} (\bibinfo{year}{2002}).

\bibitem[{\citenamefont{Sasaki et~al.}(2010)\citenamefont{Sasaki, Suzuki, and
  Saito}}]{sasaki1}
\bibinfo{author}{\bibfnamefont{K.}~\bibnamefont{Sasaki}},
  \bibinfo{author}{\bibfnamefont{N.}~\bibnamefont{Suzuki}}, \bibnamefont{and}
  \bibinfo{author}{\bibfnamefont{H.}~\bibnamefont{Saito}},
  \bibinfo{journal}{Phys. Rev. Lett.} \textbf{\bibinfo{volume}{104}},
  \bibinfo{pages}{150404} (\bibinfo{year}{2010}).

\bibitem[{\citenamefont{Pham et~al.}(2005)\citenamefont{Pham, Nore, and
  Brachet}}]{pham1}
\bibinfo{author}{\bibfnamefont{C.-T.} \bibnamefont{Pham}},
  \bibinfo{author}{\bibfnamefont{C.}~\bibnamefont{Nore}}, \bibnamefont{and}
  \bibinfo{author}{\bibfnamefont{M.-E.} \bibnamefont{Brachet}},
  \bibinfo{journal}{Physica D} \textbf{\bibinfo{volume}{210}},
  \bibinfo{pages}{203} (\bibinfo{year}{2005}).

\bibitem[{\citenamefont{Berloff and Roberts}(2001{\natexlab{b}})}]{berloff3}
\bibinfo{author}{\bibfnamefont{N.~G.} \bibnamefont{Berloff}} \bibnamefont{and}
  \bibinfo{author}{\bibfnamefont{P.~H.} \bibnamefont{Roberts}},
  \bibinfo{journal}{Journal of Physics A: Mathematical and General}
  \textbf{\bibinfo{volume}{34}}, \bibinfo{pages}{81}
  (\bibinfo{year}{2001}{\natexlab{b}}).

\bibitem[{\citenamefont{Bustamante and Nazarenko}(2015)}]{busta}
\bibinfo{author}{\bibfnamefont{M.~D.} \bibnamefont{Bustamante}}
  \bibnamefont{and}
  \bibinfo{author}{\bibfnamefont{S.}~\bibnamefont{Nazarenko}},
  \bibinfo{journal}{Phys. Rev. E} \textbf{\bibinfo{volume}{92}},
  \bibinfo{pages}{053019} (\bibinfo{year}{2015}).

\bibitem[{\citenamefont{Berloff et~al.}(2014)\citenamefont{Berloff, Brachet,
  and Proukakis}}]{berloff1}
\bibinfo{author}{\bibfnamefont{N.~G.} \bibnamefont{Berloff}},
  \bibinfo{author}{\bibfnamefont{M.}~\bibnamefont{Brachet}}, \bibnamefont{and}
  \bibinfo{author}{\bibfnamefont{N.~P.} \bibnamefont{Proukakis}},
  \bibinfo{journal}{Proc. Natl. Acad.} \textbf{\bibinfo{volume}{111}},
  \bibinfo{pages}{4675} (\bibinfo{year}{2014}).

\bibitem[{\citenamefont{Balibar}(2003)}]{balibar1}
\bibinfo{author}{\bibfnamefont{S.}~\bibnamefont{Balibar}},
  \bibinfo{journal}{S\'eminaire Poincar\'e} \textbf{\bibinfo{volume}{1}},
  \bibinfo{pages}{11} (\bibinfo{year}{2003}).

\bibitem[{\citenamefont{Berloff and Roberts}(2000{\natexlab{b}})}]{roberts3}
\bibinfo{author}{\bibfnamefont{N.~G.} \bibnamefont{Berloff}} \bibnamefont{and}
  \bibinfo{author}{\bibfnamefont{P.~H.} \bibnamefont{Roberts}},
  \bibinfo{journal}{Phys. Lett. A} \textbf{\bibinfo{volume}{274}},
  \bibinfo{pages}{69} (\bibinfo{year}{2000}{\natexlab{b}}).

\bibitem[{\citenamefont{Boue et~al.}(2011)\citenamefont{Boue, Dasgupta, Laurie,
  Lvov, Nazarenko, and Procaccia}}]{procaccia}
\bibinfo{author}{\bibfnamefont{L.}~\bibnamefont{Boue}},
  \bibinfo{author}{\bibfnamefont{R.}~\bibnamefont{Dasgupta}},
  \bibinfo{author}{\bibfnamefont{J.}~\bibnamefont{Laurie}},
  \bibinfo{author}{\bibfnamefont{V.}~\bibnamefont{Lvov}},
  \bibinfo{author}{\bibfnamefont{S.}~\bibnamefont{Nazarenko}},
  \bibnamefont{and}
  \bibinfo{author}{\bibfnamefont{I.}~\bibnamefont{Procaccia}},
  \bibinfo{journal}{Phys. Rev. B} \textbf{\bibinfo{volume}{84}},
  \bibinfo{pages}{064516} (\bibinfo{year}{2011}).

\bibitem[{\citenamefont{Pitaevskii and Stringari}(2003)}]{pitaevskii}
\bibinfo{author}{\bibfnamefont{L.}~\bibnamefont{Pitaevskii}} \bibnamefont{and}
  \bibinfo{author}{\bibfnamefont{S.}~\bibnamefont{Stringari}},
  \emph{\bibinfo{title}{Bose-Einstein condensation}}
  (\bibinfo{publisher}{Clarendon Press}, \bibinfo{address}{Oxford},
  \bibinfo{year}{2003}).

\bibitem[{\citenamefont{Donnelly}(1991)}]{donnelly2}
\bibinfo{author}{\bibfnamefont{R.~J.} \bibnamefont{Donnelly}},
  \emph{\bibinfo{title}{Quantizd Vortices in Helium II}}
  (\bibinfo{publisher}{Cambridge University Press},
  \bibinfo{address}{Cambridge}, \bibinfo{year}{1991}).

\end{thebibliography}

\end{document}